\documentclass[conference]{IEEEtran}
\IEEEoverridecommandlockouts
\usepackage{cite}
\usepackage{amsmath,amssymb,amsfonts}
\usepackage{algorithmic}
\usepackage{graphicx}
\usepackage{textcomp}
\usepackage{xcolor}
\usepackage{booktabs, tabularx}
\usepackage{subcaption}
\usepackage{algorithmic}
\usepackage[ruled, linesnumbered]{algorithm2e}
\usepackage{multirow}
\usepackage{makecell}
\usepackage{flushend}
\usepackage[braket,qm]{qcircuit}
\def\BibTeX{{\rm B\kern-.05em{\sc i\kern-.025em b}\kern-.08em
    T\kern-.1667em\lower.7ex\hbox{E}\kern-.125emX}}
\begin{document}

\bstctlcite{IEEEexample:BSTcontrol}

\title{Defending Quantum Classifiers against Adversarial Perturbations through Quantum Autoencoders}

\author{\IEEEauthorblockN{Emma Andrews, Sahan Sanjaya, and Prabhat Mishra}
\IEEEauthorblockA{
\textit{University of Florida, 
Gainesville, Florida, USA}}
}

\maketitle

\begin{abstract}
Machine learning models can learn from data samples to carry out various tasks efficiently. When data samples are adversarially manipulated, such as by insertion of carefully crafted noise, it can cause the model to make mistakes. Quantum machine learning models are also vulnerable to such adversarial attacks, especially in image classification using variational quantum classifiers. While there are promising defenses against these adversarial perturbations, such as training with adversarial samples, they face practical limitations. For example, they are not applicable in scenarios where training with adversarial samples is either not possible or can overfit the models on one type of attack. In this paper, we propose an adversarial training-free defense framework that utilizes a quantum autoencoder to purify the adversarial samples through reconstruction. Moreover, our defense framework provides a confidence metric to identify potentially adversarial samples that cannot be purified the quantum autoencoder. Extensive evaluation demonstrates that our defense framework can significantly outperform state-of-the-art in prediction accuracy (up to 68\%) under adversarial attacks.
\end{abstract}


\section{Introduction}
\begin{figure}
    \centering
    \includegraphics[width=\linewidth]{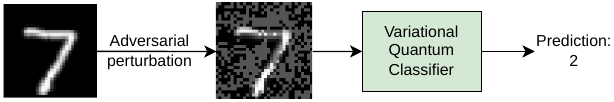}
    \caption{A clean data sample can be adversarially perturbed with noise. To humans, the image still looks like the digit 7, however, the machine learning model classifies it as a 2.}
    \label{fig:introex}
    \vspace{-0.1in}
\end{figure}

In adversarial machine learning, adversaries can craft data samples serving as inputs to machine learning models, causing the models to make incorrect decisions~\cite{qiu2019review, ren2020adversarial}. For example, Figure~\ref{fig:introex} shows that an adversary can add noise imperceptible to humans to the MNIST~\cite{lecun1998mnist} image of the digit 7, causing the model to mispredict the label as a 2. Defending against these adversarial perturbations is thus necessary to ensure a machine learning model is robust against adversarial attacks. While adversarial attacks have been well studied in classical machine learning, exploration in the context of quantum machine learning models is in its infancy.

Quantum machine learning (QML) has the potential to learn the features of data samples and carry out their tasks in less parameters compared to classical machine learning models~\cite{schuld2015introduction, lamichhane2025quantum}. QML models can still be vulnerable to adversarial data samples generated by adversarial attacks~\cite{west2023benchmarking, wendlinger2024comparative}. Specifically, adversarial attacks from classical machine learning, such as gradient-based attacks~\cite{goodfellow2015explaining, madry2019deep}, are able to produce adversarial samples that can cause a variational quantum classifier (VQC) to mispredict the class labels. Note that the model can correctly predict the class label in the absence of adversarial noise in the data samples.

There are various efforts that can successfully defend against adversarial attacks~\cite{qiu2019review, ren2020adversarial}. However, these strategies primarily rely on adversarial training where the classifier is trained on adversarial samples~\cite{wendlinger2024comparative, khatun2025classical}. There are many scenarios where this training is not feasible, such as the defense being used to protect different types of classifiers and the inability to run adversarial attacks on the classifier to generate the adversarial samples. 

To address this gap, we propose a defense framework that does not rely on adversarial training to protect against adversarial attacks. The defense framework focuses on quantum state reconstruction of the adversarial sample where the reconstruction aims to reduce or remove adversarial perturbations. Specifically, we propose a defense framework where a quantum autoencoder (QAE)~\cite{romero2017quantum} encodes the adversarial sample that would be input to the VQC. This encoded representation is decoded or reconstructed back into the original data space, effectively extracting features in the adversarial sample that were present in the training dataset. We show that  the QAE can outperform the classical autoencoder (CAE)~\cite{khatun2025classical} in prediction accuracy against adversarial attacks. 
Moreover, we provide a confidence metric from the fidelity of the encoded representation of a data sample and the classifier's class predictions to determine if the sample is a potential adversarial sample.

Specifically, this paper makes the following contributions.
\begin{itemize}
    \item We propose a framework to defend quantum classifiers against adversarial perturbations using the ability of a QAE to reconstruct the adversarial samples and purify adversarial information before it is used by VQC for classification.
    \item We provide a confidence metric to identify adversarial samples and reject a sample based on a configurable threshold. 
    \item Experimental evaluation demonstrates significant improvement (up to 68\%) in prediction accuracy compared to the state-of-the-art defense using classical autoencoders.
\end{itemize}

The rest of the paper is organized as follows. 
Section~\ref{sec:related} provides relevant background and surveys related efforts. Section~\ref{sec:method} describes our defense framework. Section~\ref{sec:results} presents the experimental results. Section~\ref{sec:conc} concludes the paper.
\section{Background and Related Work} \label{sec:related}
In this section, we first provide the necessary background on quantum machine learning, classical/quantum autoencoders, and adversarial attacks. Next, we survey related efforts and discuss their limitations.

\subsection{Quantum Machine Learning} \label{sec:bg:qml}
\begin{figure}[h]
    \centering
    \includegraphics[width=\linewidth]{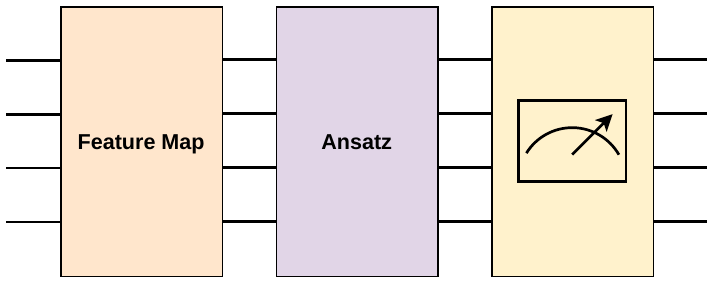}
    \caption{General structure of QML models.}
    \label{fig:qnn}
\end{figure}

Quantum Machine Learning (QML) consists of parameterized quantum circuits that can be optimized according to some objective to perform a specific functionality and achieve specific results~\cite{schuld2015introduction, biamonte2017quantum}. These models typically consist of three major components, as shown in Figure~\ref{fig:qnn}: \textit{feature map}, \textit{ansatz}, and \textit{measurements}. A \textit{feature map} embeds the input data into the quantum circuit representing the model. An \textit{ansatz} or the parameterized quantum circuit performs the computation of the model. Finally, the \textit{measurements} are used to obtain the result of the QML model. 

The parameters of the parameterized quantum circuit act on the parameterized gates, changing the exact mathematical operation that gate performs on the quantum state at a given time evolution in the parameterized quantum circuit. These parameters are optimized like the weights in classical machine learning models at each layer, changing the values to meet a specific objective. This is often done through the minimization or maximization of the objective through a loss function, such as minimizing the difference between the machine learning model class predictions and the actual class labels.

\subsection{Autoencoders} \label{sec:bg:ae}
Originating from classical machine learning, autoencoders are a group of models that can encode data samples into a smaller representation and decode those compressed representations back into the original data sample. 

\begin{figure}[h]
    \centering
    \includegraphics[width=\linewidth]{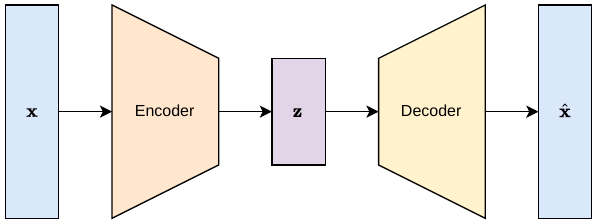}
    \caption{Classical autoencoder structure.}
    \label{fig:ae}
\end{figure}

Classical Autoencoders (CAEs) consist of two key components: an \textit{encoder} and a \textit{decoder}~\cite{hinton2006reducing}, as shown in Figure~\ref{fig:ae}. The encoder is responsible for taking an input sample $\textbf{x}\in\mathbb{R}^m$ and compressing it into a smaller dimension dataspace, known as the latent space. The latent space representation of $\textbf{x}$ is represented by $\textbf{z}\in\mathbb{R}^n,m>n$, which is given as input to the decoder to reconstruct back into the original dataspace, resulting in $\hat{\textbf{x}}\in\mathbb{R}^m$. Ideally, $\textbf{x}\approx\hat{\textbf{x}}$ for an efficient encoder and decoder.

Quantum Autoencoders (QAEs) use a similar structure to classical autoencoders, also consisting of an encoder and a decoder, to achieve data compression and reconstruction~\cite{romero2017quantum}. However, implementation and training of QAEs differ from CAEs. The encoder $U(\theta)$, parameterized by $\theta$, in the QAE is responsible for encoding the input state $|\psi_{\text{in}}\rangle$ on $n$ qubits. The encoded representation of $|\psi_{\text{in}}\rangle$ is the latent space representation, and consists of $k$ qubits, where $k<n$. The remaining $n-k$ qubits are denoted the trash qubits, and are swapped with a reference state, often $|0\rangle^{\otimes n-k}$, as input to the decoder. The decoder then uses the latent state and the reference state to reconstruct the output state $|\psi_{\text{out}}\rangle$. Due to the reversibility of quantum gates, the decoder is the inverse $U^\dagger(\theta)$ of the encoder.

\begin{figure}[h]
    \centering
    \includegraphics[width=\linewidth]{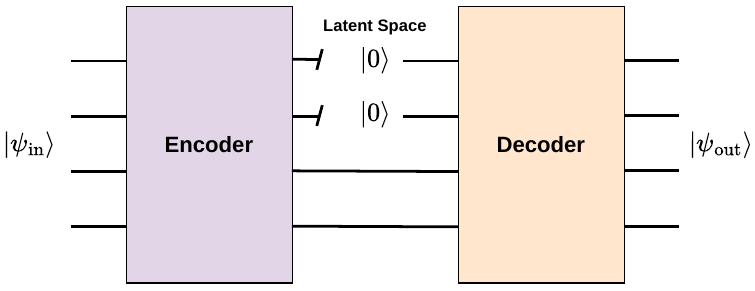}
    \caption{Quantum autoencoder structure.}
    \label{fig:qae}
\end{figure}

\subsection{Adversarial Attacks} \label{sec:bg:attack}
A machine learning model can be attacked by creating an adversarial sample where the input data sample is perturbed with specifically crafted noise to cause the machine learning model to produce an incorrect result. One major category of attacks are gradient-based attacks, where the gradients calculated during the execution of a machine learning model are used to craft an adversarial sample. The two most prominent gradient-based adversarial attacks are fast gradient sign method (FGSM)~\cite{goodfellow2015explaining} and projected gradient descent (PGD)~\cite{madry2019deep}.

\begin{figure}[htbp]
    \centering
    \includegraphics[width=\linewidth]{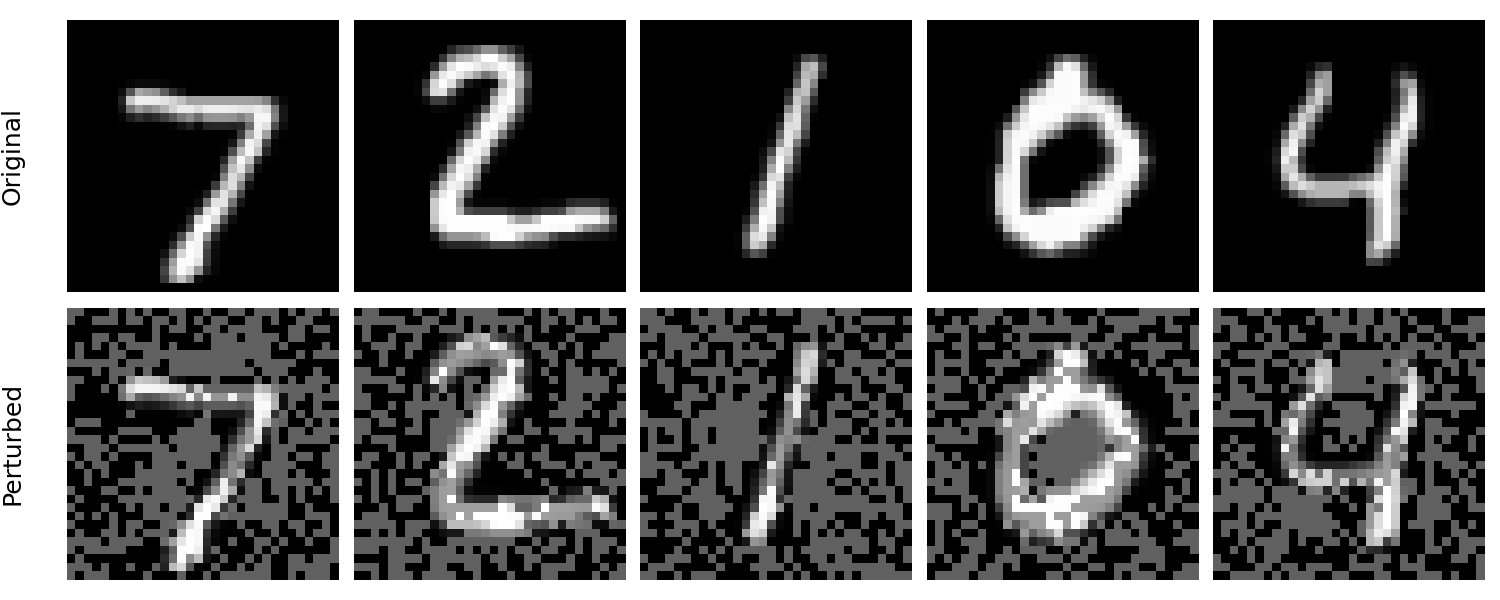}
    \caption{Example of FGSM $\epsilon=0.30$ attack on MNIST images. The original MNIST images are shown in the top row, while the adversarial images are in the bottom row.}
    \label{fig:fgsm}
\end{figure}

\begin{figure*}[htp]
    \centering
    \includegraphics[width=\linewidth]{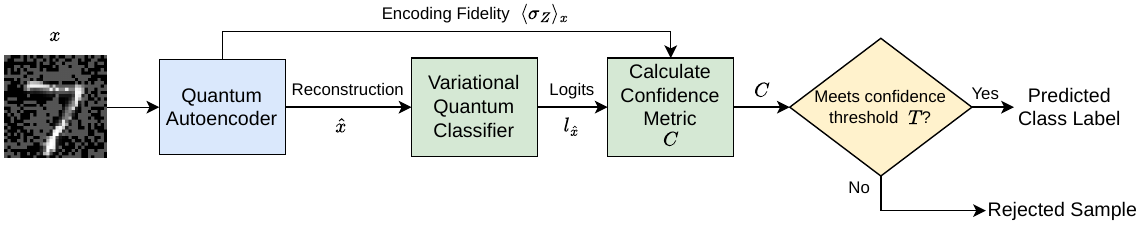}
    \caption{An overview of our defense framework. A sample, either clean or adversarial, is given as input $x$ to the QAE. The QAE produces a reconstruction $\hat{x}$ which is given to the VQC for classification. The QAE also produces the encoding fidelity $\langle\sigma_Z\rangle_x$, which is used in calculation of the confidence metric. Once the VQC has predicted the reconstruction $\hat{x}$, it produces the logit difference $l_{\hat{x}}$ which is used in addition to the encoding fidelity $\langle\sigma_Z\rangle_x$ to calculate the confidence metric $C$. This is compared against the confidence threshold $T$. If the calculated confidence metric $C$ meets the confidence threshold $T$, the predicted class label from the VQC is used. Otherwise, the sample is rejected as a potential adversarial sample.}
    \label{fig:framework}
\end{figure*}


FGSM is a popular white-box adversarial example generation method that uses the sign of the gradient of the loss function to add a fixed, small perturbation $\epsilon$ to the original input. Given an input $x$ with label $y$ and a loss function $\mathcal{L}(\cdot)$, the adversarial example $x_{\mathrm{adv}}$ is generated as
\begin{equation}
\label{eq:fgsm}
x_{\mathrm{adv}} = x + \epsilon \cdot \mathrm{sign}\!\left(\nabla_x \mathcal{L}(\theta,x, y)\right).
\end{equation}
In an untargeted attack, FGSM maximizes the loss with respect to the true label, pushing the input away from the correct classification boundary. In contrast, in a targeted attack, FGSM minimizes the loss with respect to a chosen target label by applying the perturbation in the opposite gradient direction, thereby steering the input toward an incorrect class boundary. An example of original MNIST images perturbed using FGSM with $\epsilon = 0.30$ is shown in Figure~\ref{fig:fgsm}.

PGD is another widely used white-box adversarial example generation method that extends FGSM by iteratively applying small perturbations. Given an input $x$ with label $y$ and a loss function $\mathcal{L}(\cdot)$, the adversarial example is initialized as $x^{(0)} = x$ and updated iteratively as
\begin{equation}
\label{eq:pgd}
x^{(t+1)} = \Pi_{\mathcal{B}_\epsilon(x)} \!\left( x^{(t)} + \alpha \cdot \mathrm{sign}\!\left(\nabla_x \mathcal{L}(\theta, x, y)\right) \right),
\end{equation}
where $\alpha$ is the step size and $\Pi_{\mathcal{B}_\epsilon(x)}(\cdot)$ denotes projection onto the ball of interest (e.g., the $\ell_\infty$ ball of radius $\epsilon$ centered at $x$). Unlike FGSM, which applies a single-step perturbation, PGD performs multiple constrained updates, enabling it to explore the loss landscape more effectively and generate stronger adversarial examples.

\subsection{Related Work} \label{sec:related}
There are several avenues to defend against adversarial attacks in QML. For example, QML models can be defended using noise, such as depolarization noise or noise layers~\cite{du2021quantum, huang2023enhancing}. Other defense mechanisms include adversarial training, where the defense models or classifier itself are trained on adversarial samples either in place of or in conjunction with the original, clean samples~\cite{lu2020quantum, west2023quantum}. 
One way to reinforce the adversarial training is to use a regularized loss function, such as Lipschitz regularization~\cite{berberich2024training, wendlinger2024comparative}. Lipschitz regularization enables tighter Lipschitz bounds for the trained quantum classifier. The Lipschitz bounds provide a way to quantify how input differences, such as adversarial noise, affect the output of a machine learning model. 

In classical machine learning, there are several defenses proposed using generative models to purify adversarial samples, such as autoencoders~\cite{hwang2019puvae}, generative adversarial networks~\cite{samangouei2018defensegan}, and diffusion models~\cite{nie2022diffusion}. This notion has carried over to quantum, where classical autoencoders are adversarially trained to defend VQCs against adversarial attacks through purification of the adversarial samples~\cite{khatun2025classical}. However, there are many scenarios where training with adversarial samples is either not possible or can cause overfitting to a specific attack type instead of robustness for any kind of adversarial attack~\cite{tramer2020adaptive, croce2020reliable}. Therefore, it is important to develop defenses that do not use adversarial training to successfully defend against adversarial attacks for these scenarios.

\section{Quantum Autoencoders for Defending against Adversarial Attacks} \label{sec:method}

Figure~\ref{fig:framework} provides an overview of our proposed defense framework QAE++ that consists of three major components. We first describe the VQC, which is the model under attack. Next, we describe how QAE processes the given input sample to produce a reconstructed sample. We propose a confidence metric to remove the reconstructed samples that are likely to be misclassified. Finally, we put all the components together to provide an overview of QAE++ algorithm.

\subsection{Variational Quantum Classifiers} \label{sec:method:vqc}
VQCs are a popular model architecture for classification tasks in QML~\cite{schuld2020circuitcentric}. These models embed the data into the circuit that consist of an ansatz featuring repeating layers of rotations and entanglement. The expectation values are then measured to identify the class label the VQC recognizes the input data sample as being a member of. 

An important initial step in the VQC is preparing and embedding the data into the quantum circuit for the ansatz to process. As the VQCs will be classifying images, amplitude embedding is a natural fit. Amplitude embedding takes a feature vector $x$ of size $2^n$ and embeds it into $n$ qubits, such that
\begin{equation}
    |\psi\rangle=\sum^{2^n}_{i=0}x_i|i\rangle
\end{equation}
is the resulting quantum state~\cite{schuld2018supervised}.

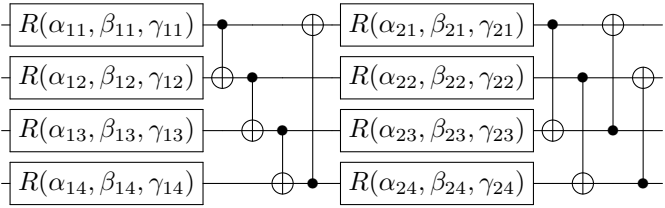
\begin{figure}
    \centering
    \centerline{\Qcircuit @C=0.32em @R=.4em {
        & \gate{R(\alpha_{11},\beta_{11},\gamma_{11})} & \ctrl{1} & \qw & \qw & \targ & \qw & \gate{R(\alpha_{21},\beta_{21},\gamma_{21})} & \ctrl{2} & \qw & \targ & \qw & \qw\\
        & \gate{R(\alpha_{12},\beta_{12},\gamma_{12})} & \targ & \ctrl{1} & \qw & \qw & \qw & \gate{R(\alpha_{22},\beta_{22},\gamma_{22})} & \qw & \ctrl{2} & \qw & \targ & \qw\\
        & \gate{R(\alpha_{13},\beta_{13},\gamma_{13})} & \qw & \targ & \ctrl{1} & \qw & \qw & \gate{R(\alpha_{23},\beta_{23},\gamma_{23})} & \targ & \qw & \ctrl{-2} & \qw & \qw \\
        & \gate{R(\alpha_{14},\beta_{14},\gamma_{14})} & \qw & \qw & \targ & \ctrl{-3} & \qw & \gate{R(\alpha_{24},\beta_{24},\gamma_{24})} & \qw & \targ & \qw & \ctrl{-2} & \qw\\
    }}
    \caption{Two layers of the circuit-centric classifier design for a four-qubit circuit. A layer consists of a group of 4 rotation gates followed by 4 entangling gates.}
    \label{fig:sel}
\end{figure}

We focus on VQCs using the layer design of circuit-centric quantum classifiers~\cite{schuld2020circuitcentric}, also known as the strongly entangling layers. A layer consists of a single qubit rotation gate $R(\phi, \theta, \omega)$ followed by CNOT gates for neighboring pairs of qubits. This repeated layering pattern is shown in Figure~\ref{fig:sel} for two layers on four qubits. Therefore, the total number of parameters of the VQC model is $n\_layers\times n\times 3$.

The ansatz must be measured to obtain the predicted class label of a given input data sample by the model. The expectation value is measured for the number of classes associated with the dataset, meaning each qubit measured represents the logit for one class. The maximum value across the resulting expectation values is the class label associated with the corresponding qubit. For example, if qubit 2 had the maximum expectation value, the predicted class label for the input data sample would be class 2.

\subsection{Quantum Autoencoders} \label{sec:method:qae}
The QAE is the central part of QAE++. It is responsible for acting as a preprocessing step on the input data samples prior to the VQC. The QAE takes the input data sample, encodes it into a latent representation, and then reconstructs the input data sample. This reconstruction emphasizes the extraction of learned features in the original, clean data samples. Thus, an adversarial sample can be reconstructed, purifying adversarial noise from the sample.

The QAE must first embed data into the circuit for the encoder to compress. This is achieved with amplitude embedding like with the VQCs as described in Section~\ref{sec:method:vqc}. The original data samples are images, where $2^n$ pixels are embedded into $n$ qubits. The encoder ansatz processes the embedded quantum state and compresses it into the latent space representation on $k$ qubits, where $k<n$. The remaining $n-k$ qubits, or the trash space, must result in the reference state to the QAE for an effective compression in the latent space. For this QAE, we use a reference state of the basis state $|0\rangle^{\otimes n-k}$.

The decoder takes the reference state and the compressed latent state as input to reconstruct the output state. This output state will encompass $n$ qubits, the same as the original input state. As the aim of the decoder is to reconstruct the state from the latent state, due to the reversible properties of quantum gates the decoder can be the Hermitian conjugate of the encoder. The result from the decoder is the reconstructed output state, which is measured with its probabilities to produce the resulting image.

To learn the features of the dataset and compress the data samples effectively, the encoder ansatz is designed with repeated layers of single-qubit rotation gates and entanglements between neighboring qubits. These layers utilize the same layer structure as the VQCs with the strongly entangling layers, with the example given in Figure~\ref{fig:sel}.

Formally, the interactions between these states through the encoder and decoder circuits can be defined as the interaction between different subsystems~\cite{romero2017quantum}. The latent space qubits are subsystem $A$, the trash space qubits are subsystem $B$, and the reference space qubits are subsystem $B'$. We can express the encoder output instead as a product of the subsystem states, as
\begin{equation}
    U(\theta)|\psi\rangle_{AB}=|\psi\rangle_A\otimes|\psi\rangle_B,
\end{equation}
where $|\psi\rangle_A$ is the state of the latent space representation, $|\psi\rangle_B$ is the trash state, and $|\psi\rangle_{AB}$ is the output state of the encoder, including both the latent state and the trash state.

The output of the decoder should be equivalent to the input of the encoder, which can be formally expressed as
\begin{equation}
    U^\dagger(\theta)U(\theta)|\psi\rangle_{AB}=|\psi\rangle_{AB}.
\end{equation}
This occurs if the trash state is the reference state, in this case, if the trash state is the basis state $|\psi_{B'}\rangle=|0\rangle^{\otimes n-k}$. Thus, the parameters $\theta$ must be optimized so that the input state is compressed into the encoder output state $|\psi_{AB}\rangle$ where $|\psi_B\rangle=|\psi_{B'}\rangle$. Therefore, the full output of the QAE is represented by the density matrix
\begin{equation} \label{eq:qae} 
    \rho_{\text{out}}=U^\dagger_{AB'}\left(\theta\right)\text{Tr}_B\left[U_{AB}\left(\theta\right)|b\rangle U_{AB'}\left(\theta\right)\right]U_{AB'}\left(\theta\right),
\end{equation}
where $|b\rangle=|\psi_{\text{in}}\rangle\langle\psi_{\text{in}}|_{AB}\otimes|a\rangle\langle a|_{B'}$, $|a\rangle$ is the reference state contained in $B'$, and $|\psi_{\text{in}}\rangle$ is the input state to the encoder.

This leads to efficient training of QAEs, where only the encoder needs to be trained to maximize the fidelity between the trash state and the reference state at the encoder output. As the decoder is the inverse of the encoder, the encoder weights can also be used for the decoder. This is unlike CAEs, where both the encoder and decoder must be trained with their own individual weights. Specifically for QAEs, the encoder can be trained to optimize the trash state resulting from the encoder to the reference state. This can be achieved by measuring the fidelity between the trash state and the reference state. 

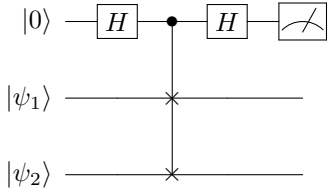
\begin{figure}
    \centering
    \centerline{\Qcircuit @C=1.2em @R=1.5em @!R {
        \lstick{\ket{0}} & \gate{H} & \ctrl{1} & \gate{H} & \meter \\
        \lstick{\ket{\psi_1}} & \qw & \qswap & \qw & \qw\\
        \lstick{\ket{\psi_2}} & \qw & \qswap \qwx[-1] & \qw & \qw
    }}
    \vspace{0.1in}
    \caption{A SWAP test used to measure the fidelity between quantum states $|\psi_1\rangle$ and $|\psi_2\rangle$.}
    \label{fig:swap}
\end{figure}

One efficient method to calculate the fidelity between two quantum states is through the SWAP test~\cite{buhrman2001quantum}. The SWAP test requires an ancilla qubit, where the Hadamard gate is performed on the ancilla before performing a controlled SWAP between the two comparing quantum states with the ancilla qubit as the control qubit. This is followed by another Hadamard gate on the ancilla qubit. The expectation value $\langle\sigma_Z\rangle$ is measured to produce the fidelity of the two quantum states. Figure~\ref{fig:swap} depicts the resulting circuit structure for the SWAP test. 

The expectation value $\langle\sigma_Z\rangle$ measured from the SWAP test representing the fidelity between the trash state and the reference state must be maximized to a value of 1 so that the trash state is equivalent to the reference state. In other words, maximization of the fidelity results in the trash state of $|0\rangle^{\otimes n-k}$ and compression of the input state $|\psi_{\text{in}}\rangle$ into $k$ qubits from $n$ original qubits. Thus, the loss function for training the encoder maximizes the fidelity with
\begin{equation} \label{eq:loss}
    \mathcal{L}=1-\langle\sigma_Z\rangle,
\end{equation}
which is minimized to 0.

\subsection{Confidence Metric} \label{sec:method:conf}

To augment the QAE reconstructions for better defense, we establish a confidence metric to determine whether to accept or reject a sample after classification by the VQC. This consists of two major components: the encoding fidelity and the logit difference. These two components are analyzed per sample to establish the confidence metric to improve overall defense.

\subsubsection{Encoding Fidelity}
A unique property with QAEs versus CAEs is that they can be trained using only the encoder by comparing the trash state $|\psi_B\rangle$ to the reference state $|\psi_{B'}\rangle$, as described previously. We utilize the SWAP test, measuring the trash state and the reference state as the encoding fidelity $\langle\sigma_Z\rangle_x$, analyzing how well the encoder is able to compress and extract the features of the adversarial sample $x$ compared to the clean dataset it was trained on. This SWAP test measures the fidelity of the compressed representation. A lower than normal fidelity compared to the original training dataset indicates that the sample contained many features that were not exhibited in the original training dataset, indicating potential adversarial samples. 

\subsubsection{Logit Difference}

When the model classifies the given data sample, it produces logits or the different class labels and its probability that the sample is of the associated class label. When doing normal classification, the maximum value in the logits and its corresponding class label is taken as the predicted class for the given data sample. With samples predicted correctly, the difference in the two maximum logits produced is large, as the model was confident in its prediction. Incorrect samples have a smaller difference, indicating that the model was not confident and predicted the class label incorrectly. This same notion extends to adversarial samples, as the logit difference will also be minimized to alter the classifier from predicting the correct class label.

Specifically, we formulate the confidence metric as a linear combination of the encoding fidelity and the logit difference. For a given sample, its calculated confidence metric is checked against a threshold confidence metric. If the calculated confidence metric is greater than the threshold, the resulting prediction is accepted, with the resulting class label being the same as the predicted label from the classifier. Otherwise, the resulting prediction is rejected as a potential adversarial sample. 

\subsubsection{Computation of Confidence Metric}
The linear combination representing the confidence metric $C$ adds the encoding fidelity with the logit difference. With these two metrics, they each have their own value ranges. While the encoding fidelity ranges from 0 to 1, the logit difference varies between 0 and 2 under expectation value measurement. To normalize the ranges, we divide the logit difference by 2 in the linear combination. This results in
\begin{equation} \label{eq:conf}
    C=\langle\sigma_Z\rangle_L+\frac{l_{\hat{x}}}{2},
\end{equation}
where the adversarial sample $x$ has encoding fidelity of $\langle\sigma_Z\rangle_x$ for its reconstruction $\hat{x}$ from the QAE with resulting logit difference of $l_{\hat{x}}$.

For a calculated confidence metric $C$, this is thresholded against an established confidence metric $T$. This threshold is expressed as the same linear combination, however the exact values may differ. While the direct values of the encoding fidelity and the logit difference are used for a calculated confidence metric, for threshold, they are determined from a validation set of samples and their resulting values. We threshold the encoding fidelities on the 1st percentile of clean validation set encoding fidelities, expressed as
\begin{equation}\label{eq:enc_fid}
    \langle\sigma_Z\rangle_T=P_1(E)-\delta,
\end{equation}
where $E$ is the resulting encoding fidelities produced from the clean validation set and $\delta$ is a small tolerance factor to account for potential outliers with clean samples.
The thresholds for the logit differences are calculated similarly, where the average logit difference of the incorrect samples from the validation set is expressed as
\begin{equation}\label{eq:log_dif}
    l_{T}=(\frac{1}{d}\sum_{i=1}^dz^i_1-z^i_2)-\gamma,
\end{equation}
where $d$ is the number of samples in the validation set, $z^i_1$ is the maximum logit for sample $i$, $z^i_2$ is the second largest logit for sample $i$, and $\gamma$ is a small tolerance factor to allow for some outliers with clean samples. The resulting threshold is thus
\begin{equation}
    T=\langle\sigma_Z\rangle_T+\frac{l_T}{2},
\end{equation}
and is used to determine if a calculated sample confidence metric $C$ is rejected or accepted.

\subsection{QAE++: Putting It All Together}
The QAE, confidence metric, and VQC are combined together into the full defense framework QAE++. Algorithm~\ref{alg:defense} outlines the major steps in our proposed framework (QAE++) shown in Figure~\ref{fig:framework}. QAE++ takes in as input a data sample to predict. This data sample can either be a clean image or an adversarial sample, which the defense framework has no prior knowledge on which one it is. In addition, the threshold $T$ for the confidence metric is obtained on clean samples from a validation dataset.

With the inputs, first the QAE++ algorithm processes the given input data sample $x$ through the QAE. The QAE reconstructs this given sample as $\hat{x}$, attempting to extract known features from the training dataset from the input sample $x$ effectively. This process also produces the encoder fidelity $\langle\sigma_Z\rangle_x$. Once the reconstruction $\hat{x}$ is obtained, it is given to the VQC for classification, resulting in the logits $z$. The difference $l_{\hat{x}}$ between the two largest logits in $z$  with the encoder fidelity $\langle\sigma_Z\rangle_x$ is used to calculate the confidence metric for the given sample, $C$ (Equation~\ref{eq:conf}). This confidence metric is compared against the given threshold $T$, resulting in either the predicted class label or a rejection as a potential adversarial sample.

\begin{algorithm}
    \caption{QAE++} \label{alg:defense}
    \SetAlgoLined
    \KwData{$x$: input data sample, either clean or adversarial, $T$: confidence metric threshold}
    \KwResult{$y$: predicted class label or rejection}
    $\hat{x}, \langle\sigma_Z\rangle_x=\text{QAE}(x)$\;
    $z=\text{VQC}(\hat{x})$\;
    $l_{\hat{x}}=z_1-z_2$\;
    $C=\langle\sigma_Z\rangle_x+\frac{l_{\hat{x}}}{2}$\;
    \eIf{$C<T$}{
        $y=\text{rejected}$\;
    }{
        $y=\text{argmax}(z)$\;
    }
\end{algorithm}
\section{Experiments} \label{sec:results}
This section evaluates the effectiveness of our defense framework against adversarial attacks. First, we outline the experimental setup. Next, we present results from several experiments to compare our framework with state-of-the-art defense approaches. 

\subsection{Experimental Setup}
We implement our defense framework with Python v3.13.5, PennyLane v0.44.0~\cite{bergholm2022pennylane}, PyTorch 2.9.1~\cite{ansel2024pytorch}, and torchvision 0.24.1. PyTorch is used to optimize and train both the quantum and classical models used throughout the experiments. 

\subsubsection{Datasets}
We use the MNIST~\cite{lecun1998mnist} and FashionMNIST (FMNIST)~\cite{xiao2017fashionmnist} datasets to evaluate our approach. Each dataset contains 60,000 training images and 10,000 testing images of handwritten digits (MNIST) or fashion items (FMNIST), where each image is grayscale of size $28\times 28$. To align with amplitude embedding, we resize each image to $32\times 32$ to fit along 10 qubits without requiring manual padding or truncating. Each image contains one item out of 10 possible items, making up the class labels.

To calculate the threshold confidence metric $T$, it must be independent of seen data in either training or testing. To achieve this, we split each dataset's testing images into 2,000 images for validation and 8,000 images for testing. The 2,000 validation images are used to gather the individual thresholds for the encoding fidelities and the logit difference, as described in Section~\ref{sec:method:conf}. The 8,000 testing images are thus used to test any defense and model.

\subsubsection{Variational Quantum Classifiers}
All VQCs are implemented in PennyLane using amplitude embedding and strongly entangling layers. These models were trained with batch size 256 and Adam~\cite{kingma2014adam} with a learning rate of 0.005 for 20 epochs. The test accuracy for the trained models is in Table~\ref{tab:all} under no attack and no defense for each dataset. For example, the MNIST VQC-100 model has a test accuracy of 81.23\% on the MNIST testing images. These VQCs are vulnerable to the FGSM and PGD adversarial attacks and will incorrectly predict class labels it previously was able to predict correctly.


\subsubsection{State-of-the-Art Classical Autoencoders}
CAEs are implemented in PyTorch using convolutional layers with ReLU activation function and a final linear layer, as used in the state-of-the-art adversarial training defense approach~\cite{khatun2025classical}. These models were trained using batch size of 128 with Adam~\cite{kingma2014adam} with learning rate of 0.001 for 10 epochs. For fair comparison, the CAEs were trained only on the clean samples of the training dataset. This is in contrast to the exact approach in~\cite{khatun2025classical} where the CAEs are trained on the adversarial samples produced from FGSM and PGD attacks.

\subsubsection{Quantum Autoencoders}
\begin{figure}[h]
    \centering
    \includegraphics[width=\linewidth]{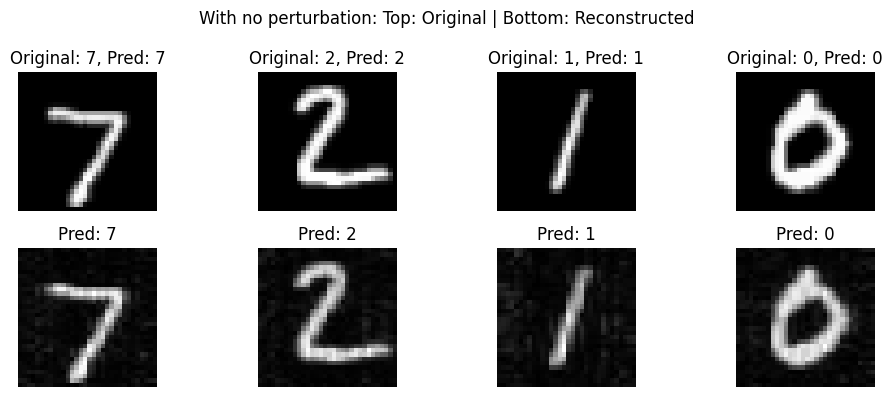}
    \caption{The MNIST VQC-100 model can still predict with about the same accuracy when the test images have been reconstructed via a QAE first. The resulting accuracy of original samples is 81.23\%, while the accuracy on QAE reconstructions of the original samples is 81.26\%.}
    \label{fig:qaeex}
\end{figure}

All QAEs are similarly implemented in PennyLane, with batch size 64, learning rate 0.1, and 10 epochs. We evaluate that each trained QAE on each dataset can effectively extract features of the corresponding dataset by evaluating the resulting VQC accuracy on reconstructed images of the original dataset testing images. MNIST QAE performs equivalently to the original test accuracy, with 81.26\% accuracy compared to 81.23\% without the QAE for VQC-100. FMNIST QAE is similar, with 64.44\% accuracy compared to the original accuracy of 65.59\% for VQC-100. The other models for each dataset and the QAE performance are given in Table~\ref{tab:all}. Figure~\ref{fig:qaeex} showcases an example of the original MNIST images and the reconstruction from the MNIST QAE on the accuracy of predictions from the MNIST VQC-100 model. The trained QAEs are used in QAE++.


\begin{table*}[htbp]
    \centering
    \caption{Individual accuracies on attack and attack power pairs per dataset, model, and defense mechanism. Best accuracies per attack pair, model, and dataset are in bold.}
    \label{tab:all}
    \begin{tabular}{lllcccccccccccc}
    \toprule
    \multirow{2}*{\textbf{Dataset}} & \multirow{2}*{\textbf{Attack}} & \multirow{2}*{\textbf{$\epsilon$}} & \multicolumn{4}{c}{\textbf{VQC-100}} & \multicolumn{4}{c}{\textbf{VQC-200}} & \multicolumn{4}{c}{\textbf{VQC-300}} \\
    \cmidrule(lr){4-7} \cmidrule(lr){8-11} \cmidrule(lr){12-15}
    & & & \textbf{None} & \textbf{CAE~\cite{khatun2025classical}} & \textbf{QAE} & \textbf{QAE++} & \textbf{None} & \textbf{CAE~\cite{khatun2025classical}} & \textbf{QAE} & \textbf{QAE++}& \textbf{None} & \textbf{CAE~\cite{khatun2025classical}} & \textbf{QAE} & \textbf{QAE++} \\
    \midrule
    \multirow{13}*{MNIST} & None & 0.00 & 81.23 & \textbf{81.47} & 81.26 & 81.26 & 84.03 & 83.79 & \textbf{84.14} & \textbf{84.14} & 85.46 & 85.31 & \textbf{85.84} & \textbf{85.84} \\
    \cmidrule(lr){2-15}
     & \multirow{6}*{FGSM} & 0.05 & 46.59 & 72.97 & \textbf{75.30} & \textbf{75.30} & 57.00 & 73.60 & \textbf{77.89} & \textbf{77.89} & 62.01 & 74.46 & 79.71 & \textbf{79.72} \\
     &  & 0.10 & 18.18 & 61.99 & \textbf{66.72} & 66.55 & 28.76 & 60.44 & \textbf{69.84} & 69.66 & 32.98 & 58.03 & \textbf{70.91} & 70.84 \\
     &  & 0.15 & 4.05 & 48.98 & 56.83 & \textbf{58.13} & 11.19 & 44.21 & 59.33 & \textbf{60.08} & 13.65 & 40.17 & 59.65 & \textbf{61.02} \\
     &  & 0.20 & 0.73 & 35.39 & 44.92 & \textbf{55.35} & 2.23 & 27.90 & 46.73 & \textbf{56.41} & 3.39 & 23.88 & 45.55 & \textbf{55.38} \\
     &  & 0.25 & 0.11 & 23.75 & 31.05 & \textbf{66.91} & 0.46 & 16.93 & 32.77 & \textbf{64.41} & 0.55 & 13.50 & 29.55 & \textbf{63.21} \\
     &  & 0.30 & 0.01 & 14.95 & 21.82 & \textbf{78.06} & 0.04 & 10.60 & 21.93 & \textbf{76.17} & 0.11 & 8.51 & 18.66 & \textbf{76.35} \\
    \cmidrule(lr){2-15}
     & \multirow{6}*{PGD} & 0.05 & 46.09 & 73.00 & \textbf{75.29} & \textbf{75.29} & 56.56 & 73.56 & 77.92 & \textbf{77.95} & 61.69 & 74.36 & 79.76 & \textbf{79.77} \\
     &  & 0.10 & 16.86 & 61.94 & \textbf{66.66} & 66.35 & 27.52 & 60.19 & \textbf{69.81} & 69.58 & 31.54 & 57.74 & \textbf{70.95} & 70.90 \\
     &  & 0.15 & 2.67 & 48.20 & 57.43 & \textbf{57.76} & 7.98 & 43.09 & 59.48 & \textbf{59.61} & 10.96 & 38.05 & 59.58 & \textbf{60.48} \\
     &  & 0.20 & 0.47 & 33.44 & 44.99 & \textbf{53.42} & 1.31 & 25.80 & 46.00 & \textbf{54.30} & 1.96 & 21.75 & 44.54 & \textbf{53.25} \\
     &  & 0.25 & 0.03 & 21.43 & 31.64 & \textbf{64.53} & 0.10 & 15.06 & 31.50 & \textbf{62.46} & 0.31 & 11.86 & 28.82 & \textbf{61.58} \\
     &  & 0.30 & 0.00 & 12.71 & 21.36 & \textbf{76.42} & 0.00 & 9.10 & 20.36 & \textbf{73.85} & 0.06 & 7.52 & 17.56 & \textbf{74.52} \\
    \midrule
    \multirow{13}*{FMNIST} & None & 0.00 & \textbf{65.59} & 64.89 & 64.44 & 64.51 & \textbf{69.26} & 68.33 & 67.89 & 67.94 & \textbf{69.10} & 68.16 & 67.51 & 67.55 \\
    \cmidrule(lr){2-15}
     & \multirow{6}*{FGSM} & 0.05 & 42.16 & \textbf{59.50} & 53.97 & 54.65 & 46.56 & \textbf{60.89} & 57.34 & 57.67 & 48.61 & \textbf{57.24} & 56.61 & 56.93 \\
     &  & 0.10 & 22.18 & \textbf{53.01} & 43.79 & 46.42 & 29.11 & \textbf{50.59} & 45.16 & 46.86 & 31.92 & 45.59 & 44.16 & \textbf{45.91} \\
     &  & 0.15 & 6.34 & \textbf{45.11} & 32.24 & 39.06 & 13.00 & \textbf{39.73} & 32.24 & 36.21 & 16.60 & 33.48 & 30.54 & \textbf{34.86} \\
     &  & 0.20 & 0.47 & \textbf{36.38} & 19.31 & 31.87 & 3.43 & 26.86 & 19.54 & \textbf{27.14} & 5.30 & 21.40 & 17.08 & \textbf{25.22} \\
     &  & 0.25 & 0.00 & 26.46 & 8.92 & \textbf{29.48} & 0.36 & 15.65 & 9.75 & \textbf{21.81} & 0.57 & 11.46 & 6.84 & \textbf{19.62} \\
     &  & 0.30 & 0.00 & 17.50 & 3.24 & \textbf{33.67} & 0.00 & 8.09 & 4.91 & \textbf{22.44} & 0.01 & 4.91 & 2.85 & \textbf{20.96} \\
    \cmidrule(lr){2-15}
     & \multirow{6}*{PGD} & 0.05 & 42.19 & \textbf{59.59} & 53.96 & 54.57 & 46.54 & \textbf{60.82} & 57.34 & 57.71 & 48.65 & \textbf{57.21} & 56.66 & 56.97 \\
     &  & 0.10 & 21.60 & \textbf{52.85} & 43.61 & 46.12 & 28.66 & \textbf{50.24} & 44.96 & 46.64 & 31.62 & 45.29 & 44.10 & \textbf{45.70} \\
     &  & 0.15 & 5.33 & \textbf{43.69} & 31.26 & 37.94 & 11.72 & \textbf{37.61} & 31.21 & 35.34 & 15.00 & 31.69 & 29.55 & \textbf{34.02} \\
     &  & 0.20 & 0.26 & \textbf{33.52} & 17.27 & 29.43 & 2.49 & 24.05 & 17.19 & \textbf{25.02} & 3.79 & 19.15 & 15.11 & \textbf{22.99} \\
     &  & 0.25 & 0.00 & 21.50 & 6.88 & \textbf{27.71} & 0.10 & 12.06 & 7.34 & \textbf{20.10} & 0.11 & 8.94 & 5.20 & \textbf{16.86} \\
     &  & 0.30 & 0.00 & 13.54 & 2.30 & \textbf{31.66} & 0.00 & 5.64 & 2.94 & \textbf{21.06} & 0.00 & 3.40 & 2.20 & \textbf{16.49} \\
    \bottomrule
    \end{tabular}
\end{table*}

\subsection{Quantum Autoencoder Reconstructions}
\begin{figure}[h]
    \centering
    \includegraphics[width=\linewidth]{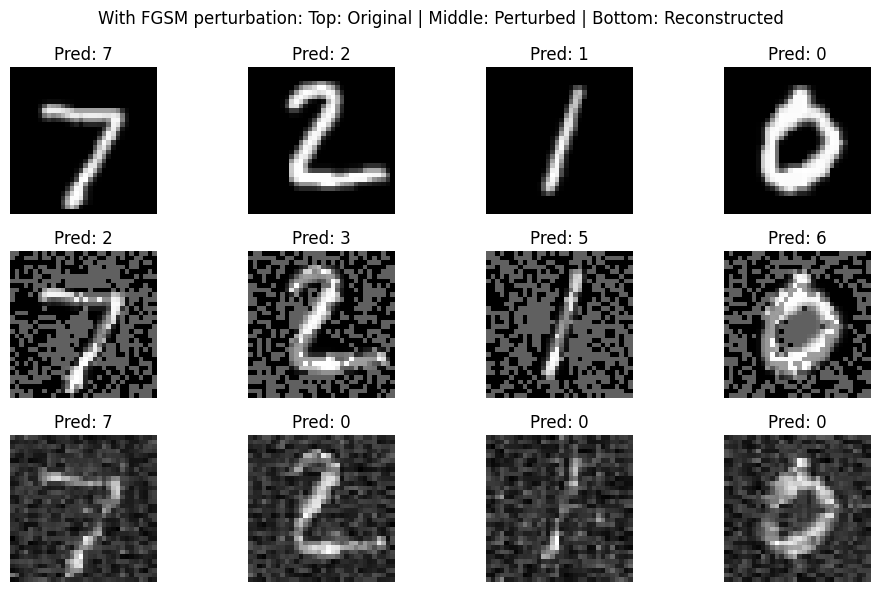}
    \caption{The MNIST images are perturbed with the FGSM $\epsilon=0.30$, as shown in the top row. The VQC prediction on these images are labeled above the images. Each image cannot be predicted correctly. The bottom row showcases the QAE reconstructions of the associated FGSM adversarial image. These reconstructions were able to make two of the images now predict correctly.}
    \label{fig:fgsmqae}
\end{figure}

\begin{figure}[h]
    \centering
    \includegraphics[width=\linewidth]{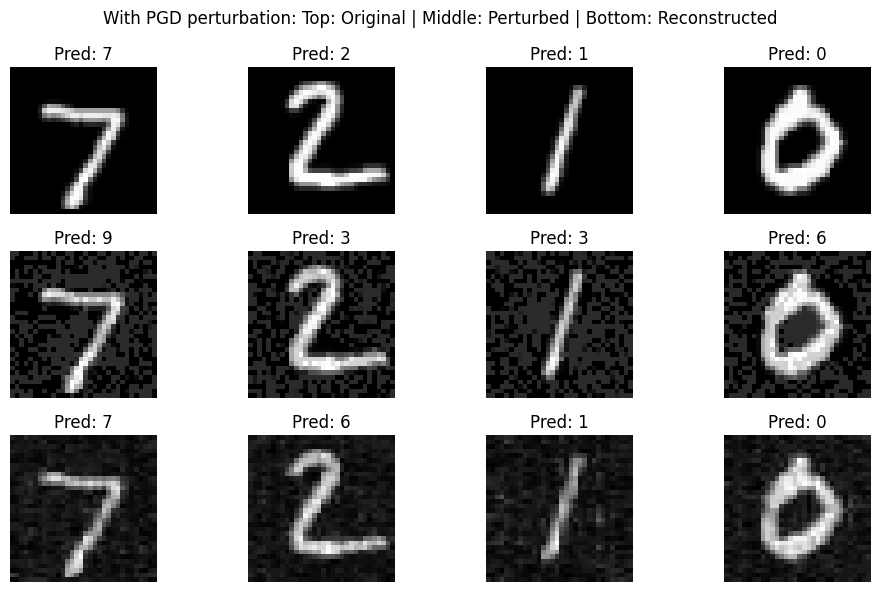}
    \caption{With the PGD adversarial perturbations, the QAE can reconstruct the images and remove some of the adversarial noise, allowing the model to correctly predict two images now than none prior without any defense.}
    \label{fig:pgdqae}
\end{figure}

The effectiveness of both the CAEs and QAEs are evaluated in the presence of adversarial samples. The testing images of the dataset are adversarially perturbed with FGSM and PGD at a specific attack power $\epsilon$, ranging from $\epsilon=0.05$ to $\epsilon=0.30$. We showcase specific examples of FGSM and PGD on the MNIST testing images and how QAEs perform. Figure~\ref{fig:fgsmqae} displays sample images perturbed with FGSM $\epsilon=0.30$ attack. The QAE can defend against this adversarial attack, allowing the VQC to classify some adversarial samples correctly whereas without a defense it was not capable of. Figure~\ref{fig:pgdqae} shows this same success for the PGD $\epsilon=0.10$ attack.

\subsection{Rejections of Samples using Confidence Metric} \label{sec:result:conf}
The confidence metric per sample is gathered from a linear combination of the encoding fidelity and the logit difference. To evaluate the effectiveness of the confidence metric and the threshold, we run the adversarial attacks individually on the entire testing image dataset to see how the confidence metric evolves under changing attack power for rejecting and accepting samples. Table~\ref{tab:ex1} displays the number of samples per possible outcome for VQC-100 for both MNIST and FMNIST, depending on if the classifier predicted the correct class label and if the confidence metric was accepted or rejected. The threshold $T$ is calculated from the results of clean samples from the validation set. The $\delta$ (Equation~\ref{eq:enc_fid}) and $\gamma$ (Equation~\ref{eq:log_dif}) values used to further threshold the threshold encoding fidelity and threshold logit difference are taken as $0.05$ and $0.01$ respectively for a tradeoff between rejecting incorrectly and correctly classified samples, as the goal is to maximize rejecting incorrect classifications and minimize correct ones. 

As the attack power $\epsilon$ increases, the number of rejected samples also increases, with the number of correctly predicted samples decreasing. This is due to the QAE producing lower quality reconstructions as the attack power grows, making it harder for the VQC to predict the class and thus decreasing the logit difference. For example with MNIST, the clean images averaged an encoder fidelity of 0.9754 and logit difference 0.1977, while for FGSM at $\epsilon=0.30$, the encoder fidelity average dropped to 0.7781 and logit difference to 0.0513. With the threshold $T$ a linear combination of the encoder fidelity and the logit difference, as these metrics decrease while the attack power increases, more samples are rejected as potential adversarial samples. Additionally, the threshold emphasizes accepting adversarial samples if the VQC is confident in its class prediction, and given its training and the QAE reconstruction, is indicative that its prediction will be of the correct class. 

This trend is observed from FGSM $\epsilon=0.05$ to $\epsilon=0.30$, where the number of rejected samples, regardless of classification outcome, increases from 2 to 7,003. The majority of the 7,003 samples rejected at $\epsilon=0.30$ are incorrectly classified, meaning the confidence metric is able to reject the most harmful samples, additionally supported by only accepting 503 incorrectly classified samples. 494 correctly classified samples are able to be accepted by the confidence metric, meaning that the QAE produced a high fidelity reconstruction that purified the adversarial noise for the VQC to classify the image correctly with a large logit difference. The resulting defense accuracies from Table~\ref{tab:all} showcase that the confidence metric rejections and acceptances is able to improve the defense accuracy from 21.82\% with the QAE only to 78.06\% with QAE++.

FMNIST has a similar phenomenon, where FGSM $\epsilon=0.30$ rejected 2,462 incorrect samples and accepted 232 correct samples. While many incorrect samples were accepted, the number of correct samples that were rejected was much smaller compared to MNIST, with only 27 correct samples being rejected. This shows that depending on the dataset and the models, the confidence metrics will result in different tradeoffs between rejection and acceptance strength, which can be configured by the user on a per dataset and model basis for best suited performance given the application usage. This rejection is still able to improve the accuracy of QAE from 3.24\% to 33.67\%, as shown in Table~\ref{tab:all}.

\begin{table}[htbp]
    \centering
    \caption{Number of rejected or accepted samples by QAE++ per classification correctness for varying attack and attack power pairs with MNIST and FMNIST VQC-100. }
    \label{tab:ex1}
    \begin{tabular}{ccccccc}
        \toprule
        \textbf{Dataset} & \textbf{Attack} & \textbf{$\epsilon$} & \makecell{\textbf{Corr.}\\\textbf{Accept}} & \makecell{\textbf{Incorr.}\\\textbf{Reject}} & \makecell{\textbf{Incorr.}\\\textbf{Accept}} & \makecell{\textbf{Corr.}\\\textbf{Reject}} \\
        \midrule
        \multirow{13}*{MNIST} & None & 0.00 & 6501 & 0 & 1499 & 0\\
        \cmidrule{2-7}
        & \multirow{6}*{FGSM} & 0.05 & 6023 & 1 & 1975 & 1\\
         && 0.10 & 5270 & 54 & 2608 & 68\\
         && 0.15 & 4204 & 446 & 3008 & 342\\
         && 0.20 & 2449 & 1979 & 2427 & 1145\\
         && 0.25 & 1108 & 4245 & 1271 & 1376\\
         && 0.30 & 494 & 5751 & 503 & 1252\\
        \cmidrule(lr){2-7}
        &\multirow{6}*{PGD} & 0.05 & 6022 & 1 & 1976 & 1 \\
         && 0.10 & 5263 & 45 & 2622 & 70\\
         && 0.15 & 4214 & 407 & 2999 & 380\\
         && 0.20 & 2482 & 1792 & 2609 & 1117 \\
         && 0.25 & 1069 & 4093 & 1376 & 1462\\
         && 0.30 & 429 & 5685 & 606 & 1280\\
         \midrule
        \multirow{13}*{FMNIST} & None & 0.00 & 5153 & 8 & 2837 & 2\\
        \cmidrule{2-7}
        & \multirow{6}*{FGSM} & 0.05 & 4315 & 57 & 3625 & 3\\
         && 0.10 & 3499 & 215 & 4282 & 4\\
         && 0.15 & 2566 & 559 & 4862 & 13\\
         && 0.20 & 1533 & 1017 & 5438 & 12\\
         && 0.25 & 699 & 1659 & 5627 & 15\\
         && 0.30 & 232 & 2462 & 5279 & 27\\
        \cmidrule(lr){2-7}
        &\multirow{6}*{PGD} & 0.05 & 4313 & 53 & 3630 & 4 \\
         && 0.10 & 3486 & 204 & 4307 & 3\\
         && 0.15 & 2496 & 539 & 4960 & 5\\
         && 0.20 & 1374 & 980 & 5638 & 8 \\
         && 0.25 & 533 & 1684 & 5766 & 17\\
         && 0.30 & 159 & 2374 & 5442 & 25\\
        \bottomrule
    \end{tabular}
\end{table}



\begin{figure*}[htbp]
    \centering
    \begin{subfigure}{0.32\linewidth}
        \centering
        \includegraphics[width=\linewidth]{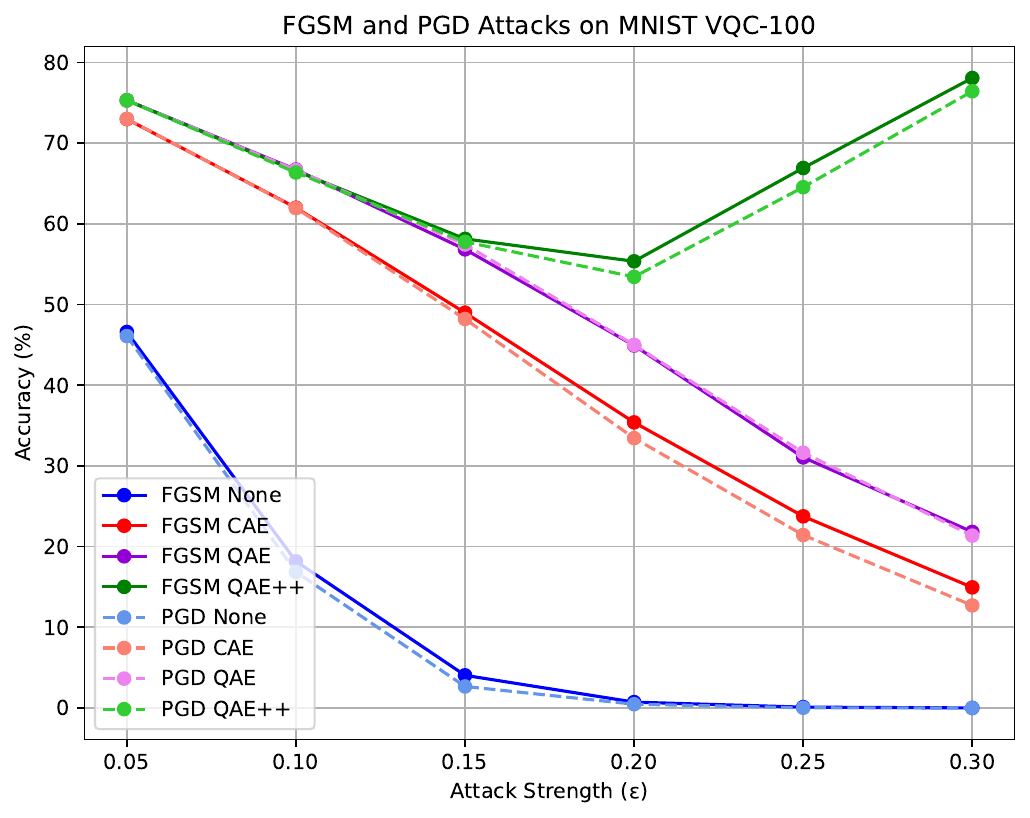}
        \caption{}
        \label{fig:mnist100}
    \end{subfigure}
    \begin{subfigure}{0.32\linewidth}
        \centering
        \includegraphics[width=\linewidth]{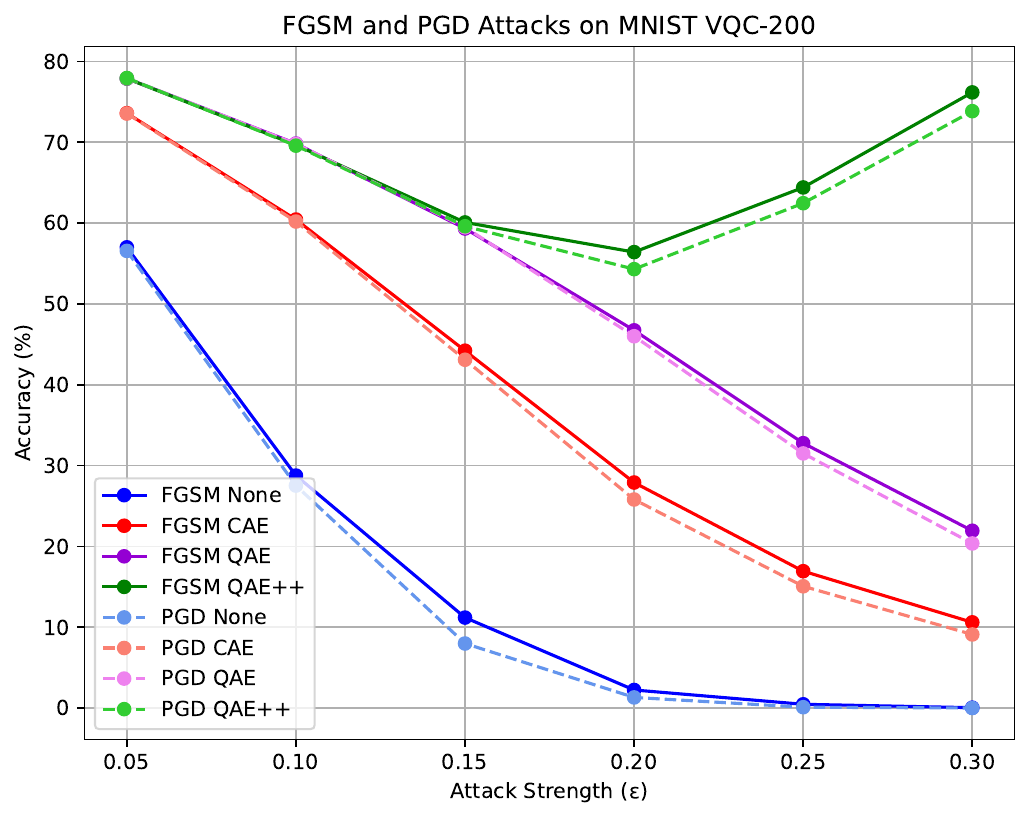}
        \caption{}
        \label{fig:mnist200}
    \end{subfigure}
    \begin{subfigure}{0.32\linewidth}
        \centering
        \includegraphics[width=\linewidth]{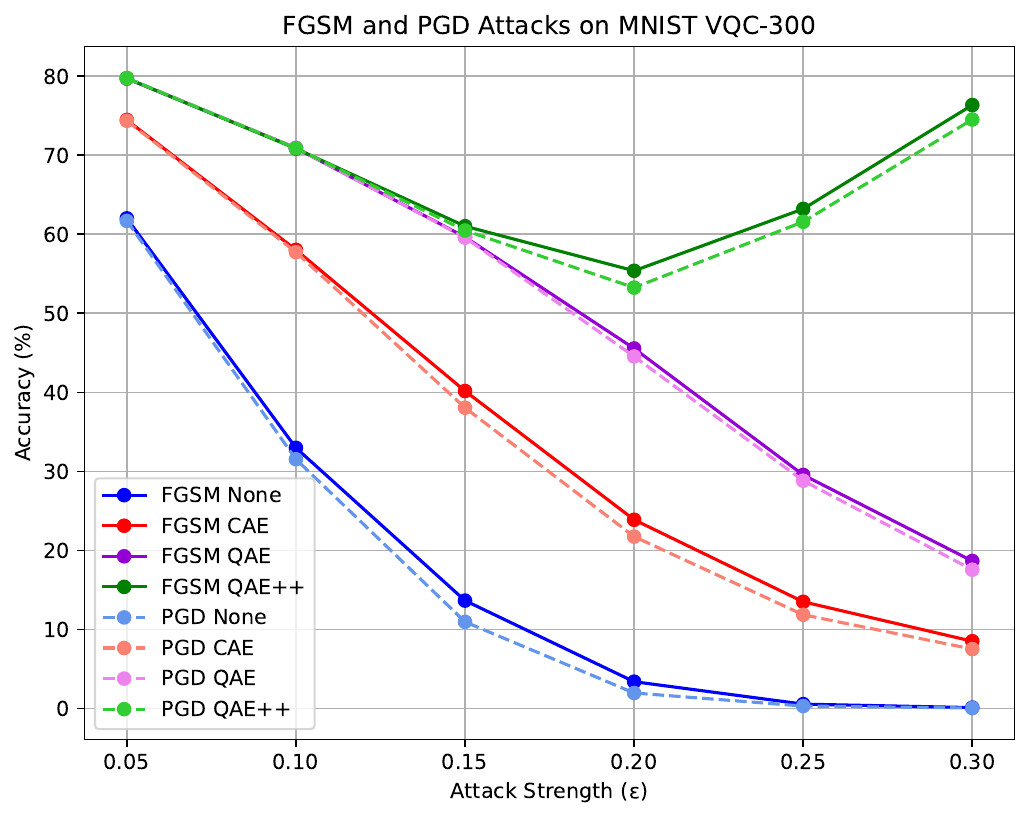}
        \caption{}
        \label{fig:mnist300}
    \end{subfigure}
    
    \begin{subfigure}{0.32\linewidth}
        \centering
        \includegraphics[width=\linewidth]{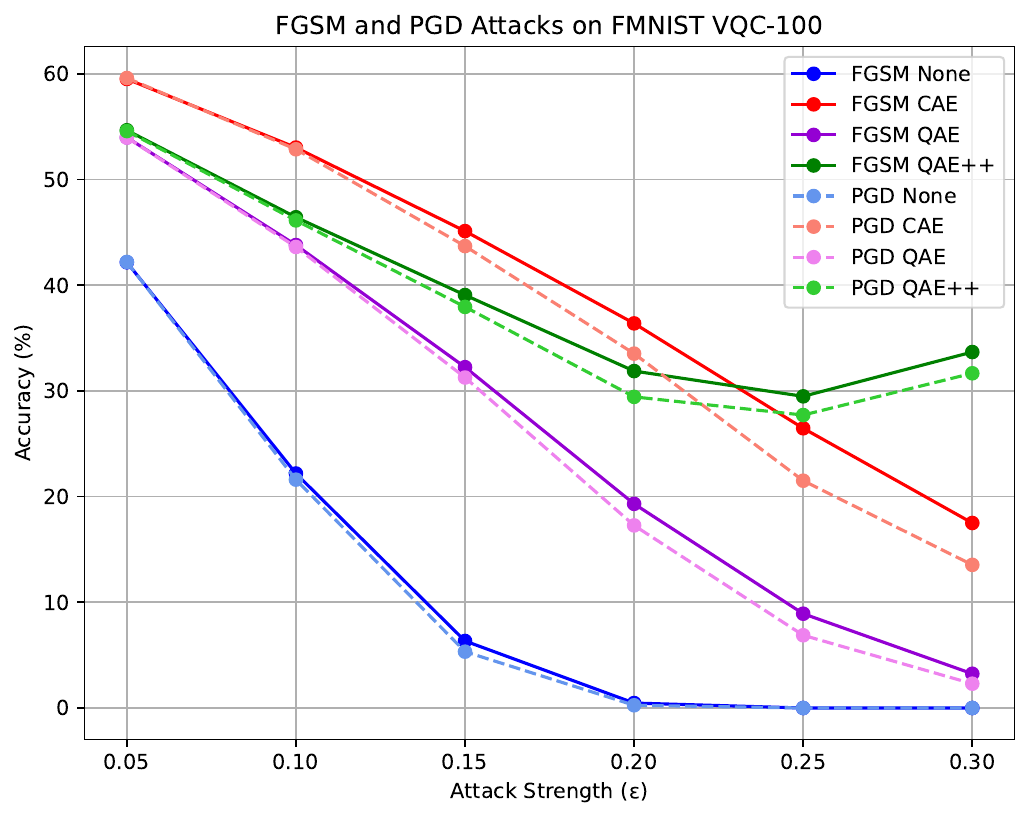}
        \caption{}
        \label{fig:fmnist100}
    \end{subfigure}
    \begin{subfigure}{0.32\linewidth}
        \centering
        \includegraphics[width=\linewidth]{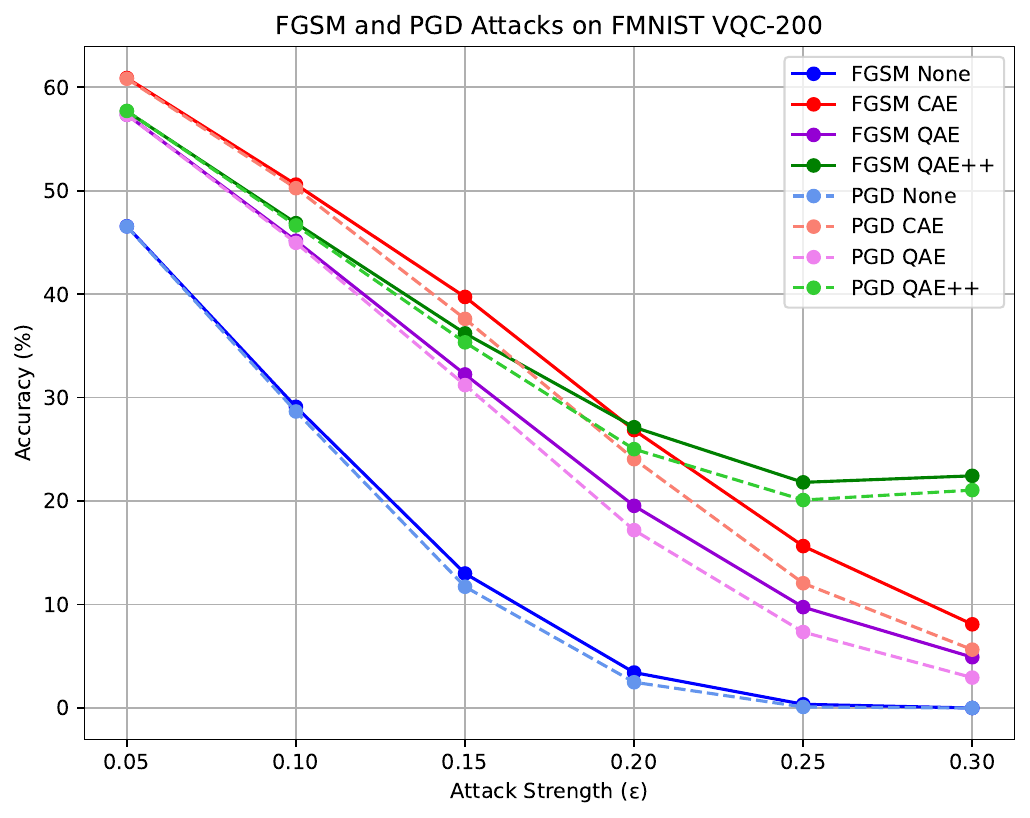}
        \caption{}
        \label{fig:fmnist200}
    \end{subfigure}
    \begin{subfigure}{0.32\linewidth}
        \centering
        \includegraphics[width=\linewidth]{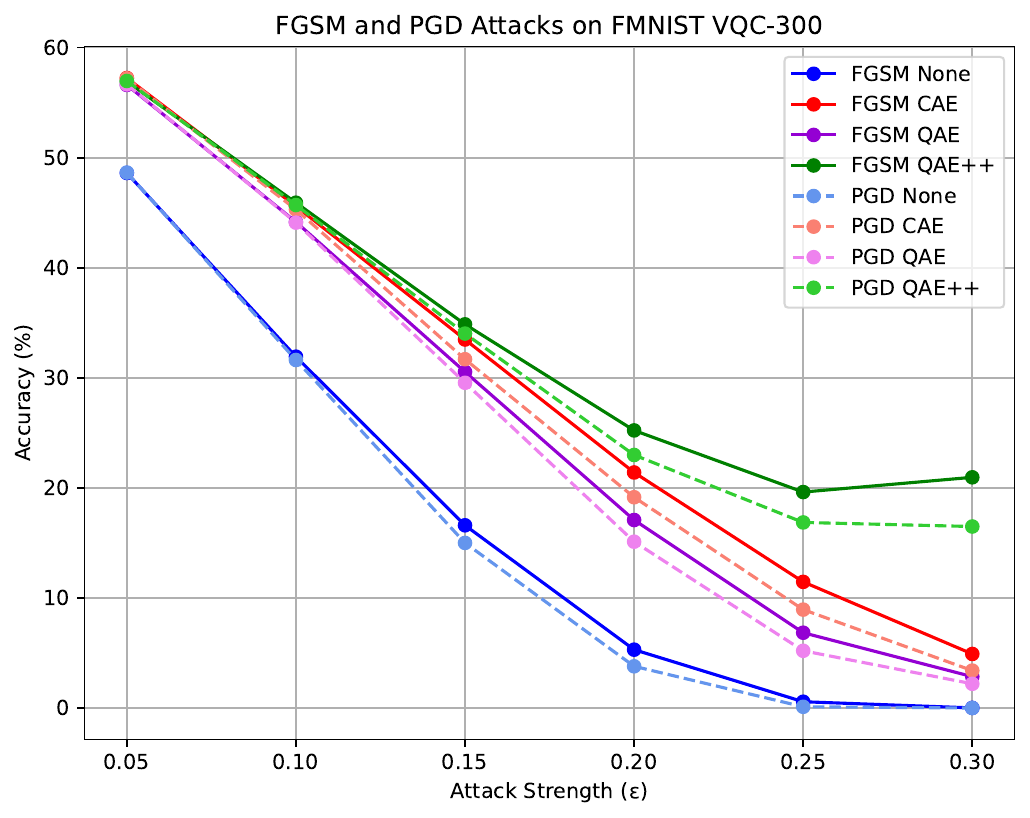}
        \caption{}
        \label{fig:fmnist300}
    \end{subfigure}
    
    \caption{Accuracy results across various models and datasets for the FGSM and PGD attacks. Each attack is defended with no defense, CAE~\cite{khatun2025classical}, QAE, or QAE++. For MNIST, across all VQCs, QAE++ is able to correctly predict and reject samples with the best accuracy. For FMNIST, as the attack power grows, QAE++ produces the best accuracy. (a) MNIST, VQC-100; (b) MNIST, VQC-200; (c) MNIST, VQC-300; (d) FMNIST, VQC-100; (e) FMNIST, VQC-200; and (f) FMNIST, VQC-300.}
    \label{fig:accuracies}
\end{figure*}

\subsection{Performance of QAE++ with Adversarial Samples}
Each attack power for each adversarial attack is tested with the 8,000 testing images in each dataset. To evaluate the performance of QAE++ on correct rejections and predictions, we utilize a final accuracy percentage. If a sample was accepted, it is counted as correct if the prediction of the VQC is of the correct class. If a sample was instead rejected, it is counted as correct if the prediction of the VQC is the incorrect class. Thus, we consider the scenarios where QAE++ accepted an incorrect prediction or rejected a correct prediction as incorrect outcomes. This can additionally be tailored given how strict the user wants the rejection to be, as for some applications, rejection of an adversarial sample regardless of the VQC prediction may be best suited. 

Figure~\ref{fig:accuracies} showcases each of the defense mechanisms for each attack pair on both the MNIST and FMNIST datasets for VQCs with 100, 200, and 300 layers. The original accuracies without any adversarial attacks are also included to establish a baseline ($\epsilon=0.00$). These accuracies are also summarized in Table~\ref{tab:all}. Across both datasets and models, as the attack power increases, QAE++ outperforms all other defense mechanisms.

With the MNIST VQC-100 results in Figure~\ref{fig:mnist100}, as the attack power grows for the FGSM attack, the accuracy of the VQC-100 model without any defense quickly becomes 0\%. Both the QAE and CAE defenses are able to retain decent accuracy under adversarial attacks, maintaining a similar accuracy up to $\epsilon=0.10$. However, at $\epsilon=0.20$, QAE++ is able to use the confidence metric to reject an increasing number of samples that the QAE and CAE struggle to reconstruct cleanly and the VQC predicts incorrectly. With FGSM of $\epsilon=0.30$, QAE++ can achieve 78.06\% accuracy by being able to reject incorrect samples and accept correct ones as shown previously in Section~\ref{sec:result:conf} and Table~\ref{tab:ex1}. In contrast, CAE and QAE are not able to reconstruct the adversarial sample cleanly enough for the VQC to correctly identify the true class label, suffering from accuracies of 14.95\% and 21.82\% respectively. For increasing attack powers, QAE++ outperforms all other defense approaches, up to 68\% better. 


This also shows that the QAE reconstructions are thus able to extract more meaningful features of the original clean images out from the adversarial sample compared to CAEs. The VQC models can use these clearer features to predict the correct label with a higher accuracy. While the QAE does not always outperform the CAE, such as for FMNIST, it is often more stable, maintaining similar resulting accuracies across different number of model layers. The reconstruction from the QAE is able to be put into a state such that information is not lost as the layers increase, which the CAE struggles with as the layers increase, resulting in a large drop in accuracy of 11.63\% versus 1.70\% for FMNIST FGSM $\epsilon=0.15$ from VQC-100 to VQC-300. The decreased accuracy from QAE can also have QAE++ perform worse compared to CAE under smaller attack powers for smaller models, as the calculated confidence metrics for these samples were not able to be sufficiently rejected in a balanced manner. An alternative configuration to reject more incorrect samples may improve accuracy, at the cost of rejecting more correct classifications.

Another benefit to using the QAE and QAE++ defense approaches is the massive decrease in the required parameters of the model compared to the CAE. The QAE model requires only 120 parameters to be optimized and stored to produce higher quality reconstructions. On the other hand, the CAE model requires about 760x more parameters, with 91,424 total to optimize and store. This is a promising approach in being able to defend VQC models without adversarial training in less parameters compared to classical model defenses, which can potentially free up computational resources for additional defense measures.

\subsection{QAE++ with Mixed Samples}
The effectiveness of the defense mechanisms are also evaluated on the testing images being split into clean and adversarial samples, instead of all images being the same type. We split the training dataset evenly into clean images and adversarial attacks of varying power. For the 8,000 testing images, they are randomly split into clean image, images perturbed with an FGSM attack, and images perturbed with a PGD attack. Within each attack, they are split into the different attack powers. We split as evenly as possible, which results in 800 clean images and 600 images for each attack and attack power. These images are given to each defense to perform their exact defense mechanism against. 

Table~\ref{tab:ex2} shows the accuracies obtained from running no defense, CAE, QAE, and QAE++ with the confidence metric and the QAE reconstruction. The confidence metric has its biggest advantage over both CAE and QAE as the attack power increases, primarily with $\epsilon=0.30$ for both FGSM and PGD. For example with MNIST VQC-100 and FGSM $\epsilon=0.30$ and its associated split of 600 images, QAE results in 152 images being classified correctly. With the confidence metric, 417 images can be correctly rejected as adversarial and 43 accepted and predicted correctly from the resulting reconstructions, leading to a boost in accuracy. This is reflected in the total difference of 51.95\% for QAE and 65.60\% for QAE++.

\begin{table}[htbp]
    \centering
    \caption{Accuracy results from mixed clean and adversarial samples of the 8,000 testing images across no defense, CAE, QAE, and QAE++. Bold values are the best accuracy for the dataset and model.}
    \label{tab:ex2}
    \begin{tabular}{cccccc}
        \toprule
        \multirow{2}*{\textbf{Dataset}} & \multirow{2}*{\textbf{Model}} & \multicolumn{4}{c}{\textbf{Accuracies}} \\
        \cmidrule(lr){3-6}
         & & \textbf{None} & \textbf{CAE~\cite{khatun2025classical}} & \textbf{QAE} & \textbf{QAE++} \\
        \midrule
        \multirow{3}*{MNIST} & VQC-100 & 17.29\% & 45.62\% & 51.95\% & \textbf{65.60\%}\\
        & VQC-200 & 21.64\% & 42.40\% & 53.69\% & \textbf{65.57\%}\\
        & VQC-300 & 24.44\% & 40.05\% & 53.02\% & \textbf{67.20\%}\\
        \midrule
        \multirow{3}*{FMNIST} & VQC-100 & 17.11\% & \textbf{41.25\%} & 29.80\% & 41.23\% \\
        & VQC-200 & 20.55\% & 36.34\% & 31.66\% & \textbf{38.75\%} \\
        & VQC-300 & 21.60\% & 31.56\% & 30.09\% & \textbf{37.21\%} \\
        \bottomrule
    \end{tabular}
\end{table}

FMNIST behaves similarly, where QAE++ performs the best out of all defenses for VQC-200 and VQC-300, while being $0.02\%$ worse than CAE for VQC-100. The confidence metric elevates the resulting accuracy by allowing reconstructions from the QAE that can cause incorrect predictions to be rejected from classification successfully. This is apparent with the difference in accuracies, especially as the layers increase, going from 30.09\% with QAE to 37.21\% with QAE++ for VQC-300. Except for VQC-100, QAE++ outperforms CAE, despite CAE outperforming QAE. Easily knowing how well the QAE performed at reconstructing the sample from its training with the encoder fidelity and the logit difference results in QAE++ being able to reject adversarial samples. 

Even though the CAE reconstructions are better quality in terms of classification compared to the QAE with 36.34\% accuracy versus 31.66\% respectively for VQC-200, being able to identify poor reconstructions via the confidence metric boosts the accuracy to 38.75\%, beyond what the CAE by itself can achieve. Despite CAE performing the best for VQC-100, it is only better by $0.02\%$. QAE++ is able to reject incorrect samples and accept correct samples, building off QAE's reconstruction ability. Thus, the increase from QAE to QAE++ from 29.80\% to 41.23\% indicates that even under poor QAE performance, QAE++ is able to improve the overall accuracy, and as QAE performs better with more VQC layers, the accuracy of QAE++ also performs better compared to CAE, as evidenced with VQC-200 and VQC-300.

\section{Conclusion} \label{sec:conc}
In this paper, we propose a defense framework QAE++, against adversarial attacks on variational quantum classifiers. QAE++ provides an adversarial training-free approach, where the defense must successfully defend against adversarial attacks without prior exposure to adversarial samples. Our QAE++ framework utilizes  quantum autoencoder that is able to purify adversarial samples and provide an encoder fidelity. This encoder fidelity is used in addition to the classifier's confidence in predicting the final class in the form of the logit difference to establish a confidence metric to identify potential adversarial samples. Through the confidence metric, samples can be rejected with incorrect classification or accepted with correct classification with high accuracy compared to state-of-the-art approaches, achieving up to 68\% better accuracy against adversarial attacks compared to state-of-the-art approaches using classical autoencoders.


\bibliographystyle{IEEEtran}
\bibliography{IEEEabrv,refs.bib}

\end{document}